# Forecasting Foreign Exchange Rate:

# A Multivariate Comparative Analysis between Traditional Econometric, Contemporary Machine Learning & Deep Learning Techniques

Manav Kaushik
Department of Economics & Finance
Birla Institute of Technology and Science, Pilani
Rajasthan, India
f2016472@pilani.bits-pilani.ac.in

A K Giri
Department of Economics & Finance
Birla Institute of Technology and Science, Pilani
Rajasthan, India
akgiri@pilai.bits-pilani.ac.in

*Abstract*—In today's global economy, accuracy in predicting macro-economic parameters such as the foreign exchange rate or at least estimating the trend correctly is of key importance for any future investment. In recent times, the use of computational intelligence-based techniques for forecasting macroeconomic variables has been proven highly successful. This paper tries to come up with a multivariate time series approach to forecast the exchange rate (USD/INR) while parallelly comparing the performance of three multivariate prediction modelling techniques: Vector Auto Regression (a Traditional Econometric Technique), Support Vector Machine (a Contemporary Machine Learning Technique), and Recurrent Neural Networks (a Contemporary Deep Learning Technique). We have used monthly historical data for several macroeconomic variables from April 1994 to December 2018 for USA and India to predict USD-INR Foreign Exchange Rate. The results clearly depict that contemporary techniques of SVM and RNN (Long Short-Term Memory) outperform the widely used traditional method of Auto Regression. The RNN model with Long Short-Term Memory (LSTM) provides the maximum accuracy (97.83%) followed by SVM Model (97.17%) and VAR Model (96.31%). At last, we present a brief analysis of the correlation and interdependencies of the variables used for forecasting.

I. INTRODUCTION

Macroeconomic and financial time series estimation is regarded as one of the most challenging applications of modern time series forecasting. The reason for this is that macroeconomic and financial time series are inherently noisy, non-stationary and chaotic as explained by [1]. While forecasting such time-series data, one general assumption made is that the past behaviour of the time-series contains all the required information which is required to predict its future behaviour. Thus, traditionally most of the attempts made to forecast foreign exchange rate focused majorly only on univariate time series analysis [2, 3, 4] popularly employing approaches like Auto-Regressive Integrated Moving Average (ARIMA) and Random Walk (RW). Because of its popularity for the past several decades, the ARIMA model [5] has been used as a benchmark to evaluate several new modelling approaches [6]. However, the major problem with ARIMA is that it is a general univariate model and is developed based on two major assumptions: (i) the time series being forecasted is linear and (ii) the time series being forecasted is stationary [7]. Moreover, univariate time series models fail to take into account the effects of other parameters which might be crucial while determining the future value of a specific macroeconomic variable.

Through this paper, we try to highlight the effectiveness of multivariate time series forecasting. As stated by [8], multivariate models can rely on greater information, where not only the lagged time series is being forecasted but also other indicators (such as technical, fundamental, inter-marker etc. for financial market), are combined to act as independent predictors. We present three very different techniques and compare their performance. The first model that we implement is one of the most popular extensions of the ARIMA model: Vector Auto-Regressive (VAR) model which is traditionally considered as a benchmark for multivariate time series analysis and forecasting [9]. However, the VAR model usually fails at mapping a nonlinear association between different variables and thus, have poor generalization. Moreover, like the ARIMA model, the VAR model also requires the input data series to be stationary.

The second model that we implement and analyze is the Support Vector Machine. Recently, Support Vector Machines developed by [10] have provided another novel approach to improve the generalization property for multivariate forecasting and prediction. SVMs adopt a Structural Risk Minimization approach, which seeks to minimize an upper bound of the generalization error rather than minimize the training error unlike most of the traditional learning techniques that adopt the Empirical Risk Minimisation Principle [11]. This results in better generalization than conventional techniques. But SVMs usually underperform if the training dataset is large or too noisy.

In recent years, Neural Network assisted multivariate analysis has become a dominant and popular tool for time series forecasting. A neural network is much more effective in mapping the dynamics of non-stationary time-series given its unique non-parametric, non-assumable, noise-tolerant and adaptive properties. Neural networks are well-known function approximators that can map any nonlinear function without any prior assumptions about the data. Recently, Artificial Neural Networks (ANNs) are being widely used to map nonlinear relationships between macroeconomic time series [12]. However, such ANN-based Multi-Layer Perceptron (MLP) models very often face the problem of overfitting, backpropagated error decay, and it cannot automatically determine the optimal time lags while fitting time-series data [13]. Thus, here we present a Recurrent Neural Network approach using Long Short-Term Memory (LSTM) which can capture the nonlinearity and randomness of time series data more effectively, as well as overcome the problem of back-propagated error decay through memory blocks of LSTM, and thus shows superior capabilities for time series prediction with long temporal dependency. With the ability to memorize long historical data and automatically determine the optimal time lags, the LSTM RNN achieves higher prediction accuracy and generalizes well with different prediction intervals [13].

According to our analysis, the LSTM RNN Model gave the performance with minimum error followed by the SVM model and VAR model.

## II. REVIEW OF THE LITERATURE

In terms of published work, the forecasting research literature is rich in recent times mainly due to the development of information technology. From the experimental results it is evident that as opposed to the traditional statistical and econometric models such as ARIMA (Auto-Regressive Integrated Moving Average), and VAR (Vector Auto Regression), neural network models produce superior results, demonstrating their suitability for forecasting the foreign exchange rates. Both ARIMA time series model and neural networks were explored by [14] for Turkish TL/US dollar exchange rate series. Most of the research work done in the literature has followed the method of univariate time series forecasting [2, 3, 4] assuming that all the information required to predict the future exchange rate is contained in the past values of the exchange rate.

Research is also done to measure and compare the performance of stochastic, ANN, SVR models in predicting the day-to-day exchange rates. [15].

Support Vector Machines developed by [10] have provided another novel approach to improve the generalization property for multivariate forecasting and prediction. SVMs are better at generalization than conventional techniques they usually underperform if the training dataset is large or too noisy.

In several applications, [16, 17, 18, 19], and several other research works have shown that ANNs perform better than ARIMA models, specifically, for more irregular series and for multiple-period-ahead forecasting. [20] provided a general introduction of how a neural network model should be developed to model financial and economic time series. However, ANN-based Multi-Layer Perceptron (MLP) models very often face the problem of overfitting, backpropagated error decay, and it cannot automatically determine the optimal time lags while fitting time-series data [13].

In recent literature, Long Shor-Term Memory Recurrent Neural Networks have been gaining popularity. Especially for the wide array of problems pertaining to time series forecasting, LSTM RNNs have shown commendable performance [21, 22]. Moreover, research works like [23] suggest how LSTM RNNs are superior to traditional econometric and statistical models such as ARIMA.

## III. METHODOLOGY & DATA

### A. Data

We have used monthly historical data from April 1994 to December 2018 for the United States and India to develop different models that can estimate USD-INR Foreign Exchange Rate. We used Bloomberg and Federal Reserve Economic Data for our analysis. Based on the research done in the literature [24, 25, 26], we selected some of the most influential macroeconomic variables which can help the fluctuation in exchange rates and thus, make better predictions using a multivariate approach. The following is the list of variables used:

- Consumer Price Index (CPI)
- Index of Industrial Production (IIP)
- Interest Rates
- Money Supply
- Total Reserves
- Stock Market Index
- Trade (Net Exports)

### B. Exploratory Data Analysis & Preprocessing

- *Visualization:*

    The following charts depict the trends and movements, for the time frame taken into consideration, of all the variables taken in our analysis.

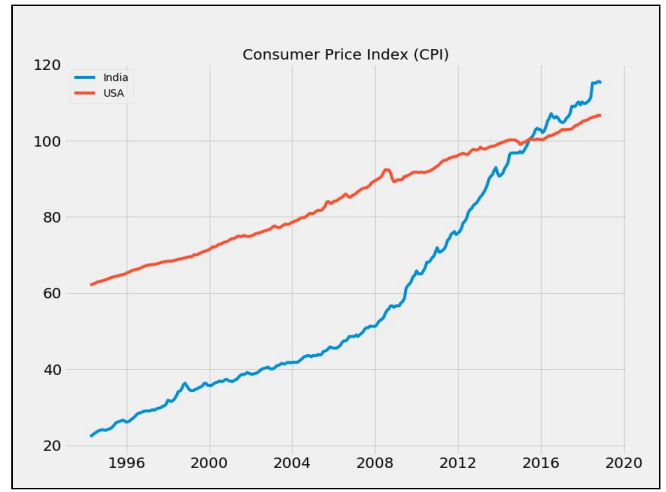

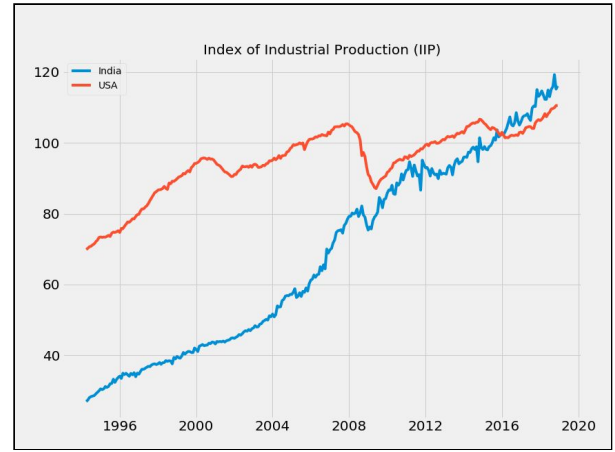

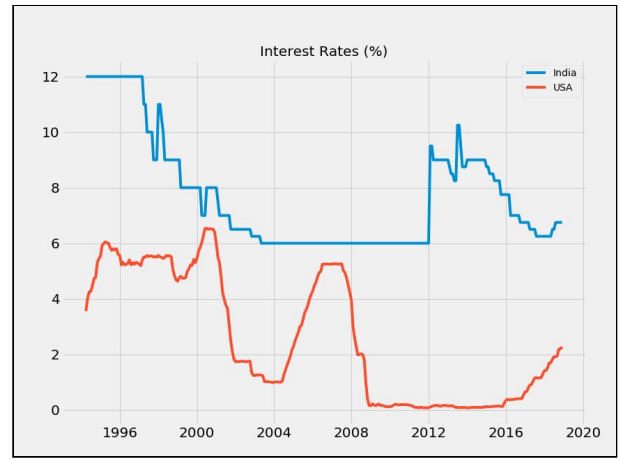

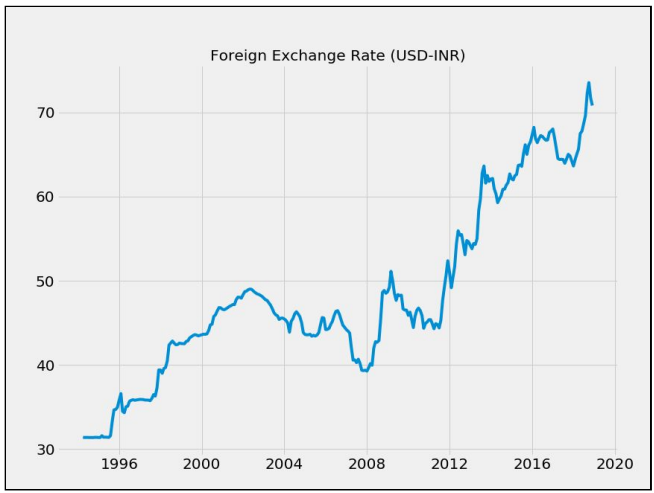

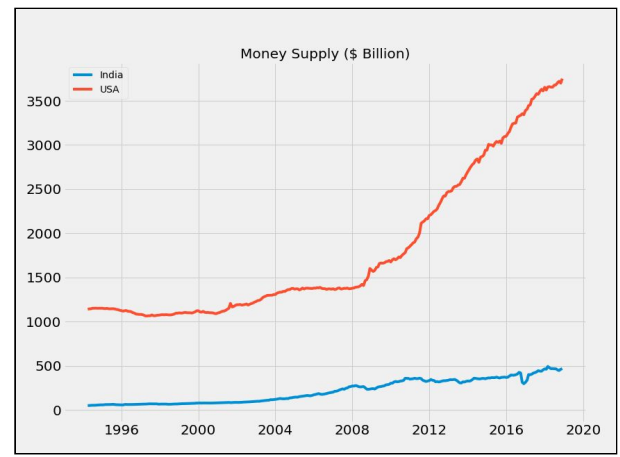

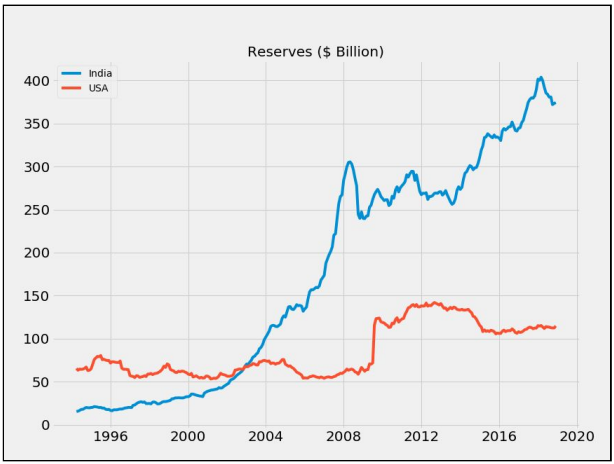

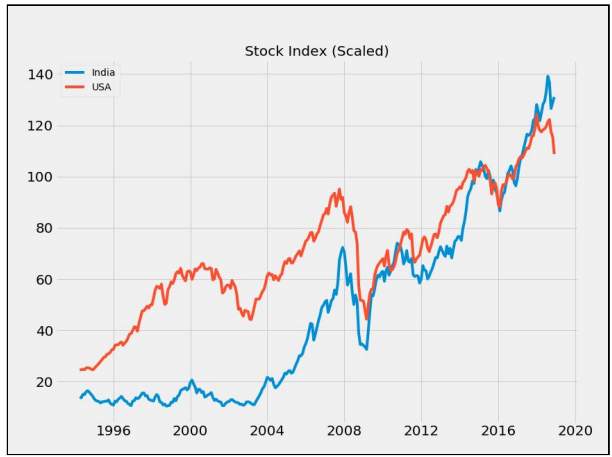

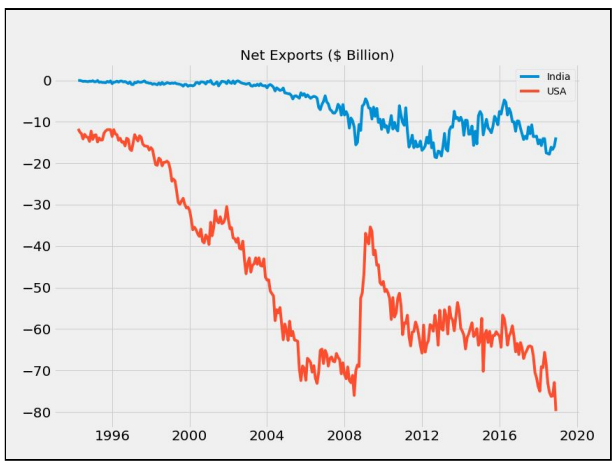

- *Input Data Scaling:*

To forecast the USD-INR exchange rate the input used for our models were the difference in the values of each macroeconomic variable of the respective country.

$$X_i(t) = X_i(USA, t) - X_i(IND, t)$$

where $X_i(USA, t), X_i(IND, t)$ are values of a macroeconomic parameter for USA and India respectively at a certain time *t*.

After this, the inputs were scaled down in the range of 0 to 1 so as to feed the model with normalized data for better convergence.

$$X_{input}(t) = \frac{(X_i(t) - X_{i,min})}{(X_{i,max} - X_{i,min})}$$

- *Granger's Causality Test:*

The Granger causality test is a method of statistical hypothesis testing performed to determine whether a time-series is useful in forecasting another. Using Granger's Causality Test, it's possible to obtain any relationship between different time-series before even building the model [27, 28].

Multivariate Granger causality analysis is generally done by fitting a Vector Auto-Regressive (VAR) model to the time-series data. Mathematically, let $X(t) \in \Re^{m \times 1} \forall \ t = 1, 2, ...., T$ be an *m*-dimensional multivariate time series. Granger causality is performed by fitting a VAR model with *L* time lags as follows:

$$X(t) = \sum_{\tau=1}^{L} A_\tau X(t - \tau) + \varepsilon(t)$$

where *t* is a white Gaussian random vector, and $A_\tau$ is a matrix for every $\tau$. A time-series $X_i$ is called a Granger cause of another time series $X_j$ if at least one of the elements $A_\tau(j, i)$ for $\tau = 1, 2, ...., L$ is significantly larger than zero (in absolute value). Granger's causality tests the null hypothesis that the coefficients of past values in the regression equation is zero i.e. the past values of time series $X_i$ do not cause the other series $X_j$. So, if the p-value obtained from the test is lesser than the significance level of 0.05, then, we can safely reject the null hypothesis.

From the results of the test, it is evident that all the variables involved are causing some variation in the foreign exchange rate.

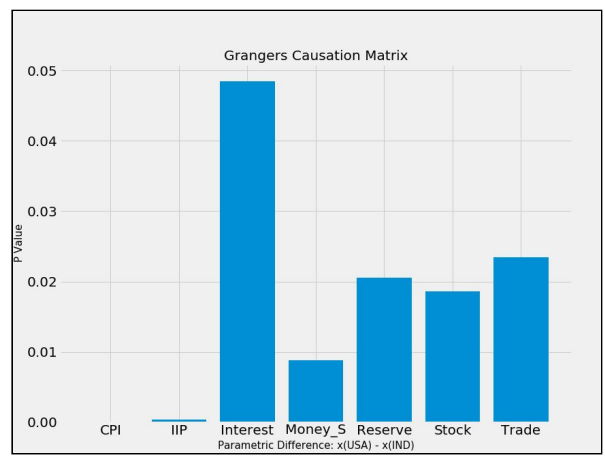

- *Augmented Dickey-Fuller Test:*

In econometrics, Augmented Dickey-Fuller (ADF) is a technique of testing the null hypothesis that a unit root is present in a time series sample. Presence of unit root implies that the series is non-stationary. The alternative hypothesis is that the series under test is stationarity or trend-stationarity. It is an augmented version of the Dickey-Fuller test for a larger and more complicated set of time series models.

The ADF statistic, used in the test, is a negative number. The more negative it is, the stronger the rejection of the hypothesis that there is a unit root at some level of confidence. If a series is found to be non-stationary, you make it stationary by differencing the series once and repeat the test again until it becomes stationary.

The ADF test is carried out in the following procedure:

$$\Delta y_t = \alpha + \beta t + \gamma \Delta y_{t-1} + \delta_1 \Delta y_{t-1} + \ldots + \delta_{p-1} \Delta y_{t-p+1} + \varepsilon_t$$

where $\alpha$ is a constant, $\beta$ is the coefficient on a time trend and $p$ is the lag order.

The unit root test is carried out with the null hypothesis $\gamma = 0$, against the alternative hypothesis $\gamma < 0$. The test statistic is calculated as:

$$DF_\tau = \frac{\hat{\gamma}}{SE(\hat{\gamma})}$$

which is then compared with the critical value at the selected significance level. Here, we have chosen a significance level of 5% which is the most commonly used level in the literature.

Following are the results of the ADF test we conducted.

| Series | Test Statistic | Critical Value at 5% significance | Remark | Stationary / Non-Stationary |
|---|---|---|---|---|
| **Forex** | -0.526 | -2.873 | Null Hypothesis cannot Be Rejected | Non-Stationary |
| **Forex (differenced once)** | -6.077 | -2.873 | Null Hypothesis is is Rejected | Stationary |
| **CPI** | 1.030 | -2.873 | Null Hypothesis is cannot Be Rejected | Non-Stationary |
| **CPI (differenced once)** | -1.926 | -2.873 | Null Hypothesis is cannot Be Rejected | Non-Stationary |
| **IIP** | 0.008 | -2.873 | Null Hypothesis is cannot Be Rejected | Non-Stationary |
| **IIP (differenced once)** | -3.173 | -2.873 | Null Hypothesis is is Rejected | Stationary |
| **Interest** | -1.638 | -2.873 | Null Hypothesis is cannot Be Rejected | Non-Stationary |
| **Interest (differenced once)** | -14.029 | -2.873 | Null Hypothesis is is Rejected | Stationary |
| **Money Supply** | 1.853 | -2.873 | Null Hypothesis is cannot Be Rejected | Non-Stationary |
| **Money Supply (differenced once)** | -2.946 | -2.873 | Null Hypothesis is is Rejected | Stationary |
| **Reserves** | -0.353 | -2.873 | Null Hypothesis is cannot Be Rejected | Non-Stationary |
| **Reserves (differenced once)** | -6.152 | -2.873 | Null Hypothesis is is Rejected | Stationary |
| **Stock Index** | -0.966 | -2.873 | Null Hypothesis is cannot Be Rejected | Non-Stationary |
| Stock Index (differenced once) | -12.417 | -2.873 | Null Hypothesis is Rejected | Stationary |
| Net Exports | -1.565 | -2.873 | Null Hypothesis cannot Be Rejected | Non-Stationary |
| Trade (differenced once) | -5.003 | -2.873 | Null Hypothesis is Rejected | Stationary |

ADF test is a necessary test to ensure and gain stationarity in the data as the VAR model is applicable only for stationary time series.

## C. Vector Auto-Regressive (VAR) Model

Vector Auto Regression (VAR) is a standard technique used in macroeconomics and is widely used for structural analysis and time series forecasting. In the VAR model, each variable is modelled as a linear combination of past values of itself and the past values of other variables in the system. For multiple time series influencing each other, it is modelled as a system of equations with one equation per variable. Since we are only concerned about forecasting foreign exchange rate, we shall only regress forex time series over its past values and past values of other variables.

As seen from the ADF test for stationarity, all the series except CPI are stationary when differenced once. Thus, the differenced time series shall be the input to the model. Moreover, considering the Akaike's Information Criterion (AIC), we select the lag order of 3 for our VAR model [29].

The formulated model looks as follow:

$$Y_{forex,\,t} = \alpha + (\beta_{1,1} Y_{forex,\,t-1} + \beta_{1,2} Y_{forex,\,t-2} + \beta_{1,3} Y_{forex,\,t-3})$$

$$+ (\beta_{2,1} Y_{cpi,\,t-1} + \beta_{2,2} Y_{cpi,\,t-2} \ldots\ldots\ldots + \beta_{8,3} Y_{trade,\,t-3}) + \varepsilon_t$$

where $Y_{i,\,t}$ represents the first difference of the series $i$ at time $t$, $\beta$ represents their respective coefficient and $\varepsilon_t$ is the error term at time $t$.

To ensure that there is no Autocorrelation in our input series, we carried out the Durbin Watson Test which is a statistical test for autocorrelation in a data set. The DW statistic always has a value between zero and 4.0. A value of 2.0 indicates that there is no autocorrelation detected in the sample. A value from zero to 2.0 indicates positive autocorrelation while value from 2.0 to 4.0 indicates negative autocorrelation.

The DW statistic showed that there is no serial correlation in our input series, which is differenced once, as all the statistic values lied very close to 2.0. While the initial series (without differencing) faced serious autocorrelation

| Variables (differenced once) | DW Statistic Value |
|---|---|
| Forex | 1.99 |
| CPI | 1.96 |
| IIP | 2.01 |
| Interest Rate | 1.98 |
| Money Supply | 1.99 |
| Reserves | 2.04 |
| Stock Index | 1.97 |
| Trade | 1.99 |

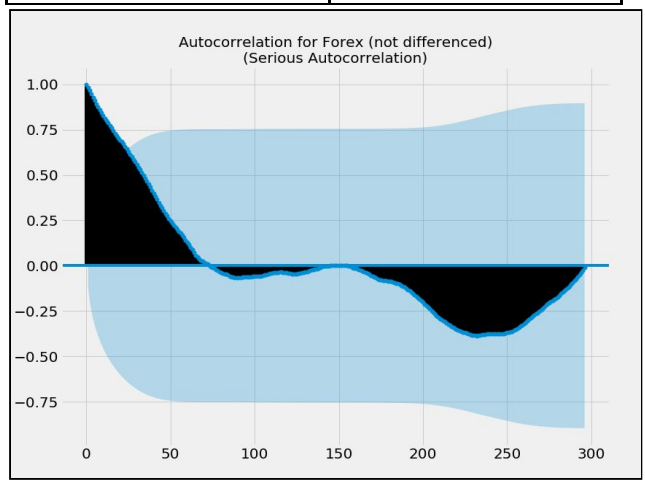

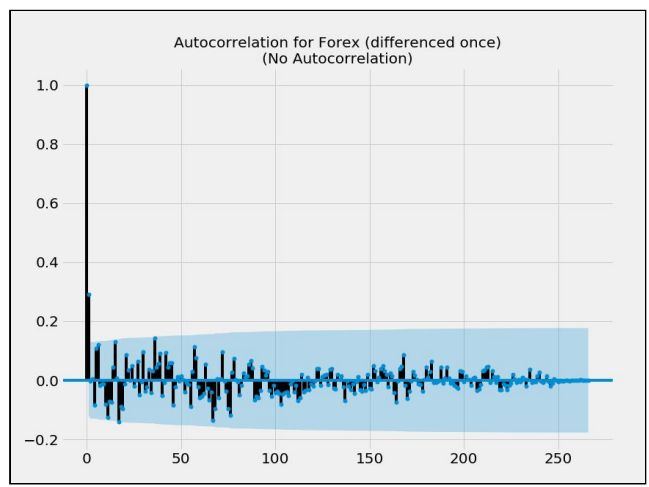

### D. Support Vector Machine (SVM) Model

Support vector machine (SVM) analysis is a popular machine learning tool for classification and regression, first identified by Vladimir Vapnik and his colleagues in 1992 [10]. SVM regression is considered a nonparametric technique because it relies on kernel functions.

Statistics and Machine Learning linear epsilon-insensitive SVM (ε-SVM) regression, which is also known as L1 loss. In ε-SVM regression [30], the set of training data includes predictor variables and observed response values. The goal is to find a function $f(x)$ that deviates from $y_n$ by value no greater than ε for each training point $x$, and at the same time is as flat as possible. To obtain the optimal value for the problem, a Lagrangian function is constructed and is solved subject to the Karush-Kuhn-Tucker (KKT) complementarity conditions.

We used the sklearn library for building our SVM Regression model and to develop a forecast for our time series. The hyperparameters' value decided after running a GridSearch on the model. The best results were obtained with regularization parameter $C$ set at 1000, kernel type as rbf and kernel coefficient, gamma, as 0.001.

### E. Long Short-Term Memory (LSTM) RNN Model

Long Short Term Memory networks are a special kind of RNN, capable of learning long-term dependencies. They were introduced by Hochreiter & Schmidhuber (1997) [31], and they work tremendously well on a large variety of problems and are now widely used.

LSTMs are explicitly designed to avoid the long-term dependency problem. Remembering information for long periods of time is their default behaviour, not something they struggle to learn.

All recurrent neural networks have the form of a chain of repeating modules of the neural network. In standard RNNs, this repeating module will have a very simple structure, such as a single *tanh* layer. LSTMs also have this chain-like structure, but the repeating module has a different structure. Instead of having a single neural network layer, there are four, interacting in a very special way.

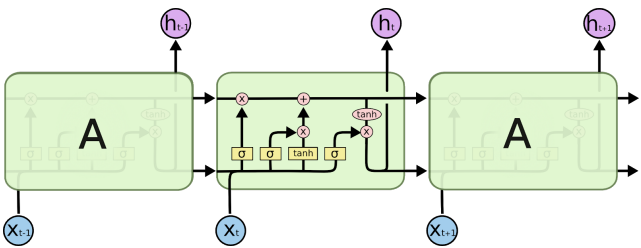

The LSTM RNN architecture is composed of one input layer, one recurrent hidden layer whose basic unit is memory block instead of traditional neuron node, and one output layer. Memory blocks are a set of recurrently connected subnets. Each block contains one or more self-connected memory cells and three multiplicative units: the input, output and forget gates, which provide continuous analogues of write, read and reset operations on the cells. The multiplicative gates allow LSTM memory cells to store and access information over long periods of time, thereby mitigating the vanishing gradient problem. For example, as long as the input gate remains closed, the activation of the cell will not be overwritten by the new inputs arriving in the network, and can, therefore, be made available to the net much later in the sequences, by opening the output gate.

For our model, the inputs given to predict a target value at time *t* are the forex vale at time *t-1* and the difference of all the macroeconomic variables at time *t-1* i.e.

$$X_i(t) = X_i(USA, t-1) - X_i(IND, t-1)$$

Suppose the input historical macroeconomic data is denoted as:
$x = (x_1, x_2, ....., x_T)$
The LSTM RNN computes the hidden vector sequence:
$h = (h_1, h_2, ....., h_T)$
And the output predicted sequence:
$y = (y_1, y_2, ....., y_T)$

By iterating the following equations:
$$h_t = H(W_{xh}x_t + W_{hh}h_{t-1} + b_h)$$
$$y_t = W_{hy}h_t + b_y$$

where the *W* term denotes the weight matrices, *b* term denotes the bias vectors and *H* term denotes the hidden layer function, which is implemented by the following composite function:
$$i_t = \sigma(W_{xi}x_t + W_{hi}h_{t-1} + W_{ci}c_{t-1} + b_i)$$
$$f_t = \sigma(W_{xf}x_t + W_{hf}h_{t-1} + W_{cf}c_{t-1} + b_f)$$
$$c_t = f_t c_{t-1} + i_{tg}(W_{xc}x_t + W_{hc}h_{t-1} + b_c)$$
$$o_t = \sigma(W_{xo}x_t + W_{ho}h_{t-1} + W_{co}c_{t-1} + b_o)$$
$$h_t = o_t h(c_t)$$

where,
$\sigma(x) = \frac{1}{1+e^{-x}}$
$g(x) = \frac{4}{1+e^{-x}} - 2$
$h(x) = \frac{2}{1+e^{-x}} - 1$

The following figure provides an illustration of the LSTM RNN prediction model architecture with one memory block.

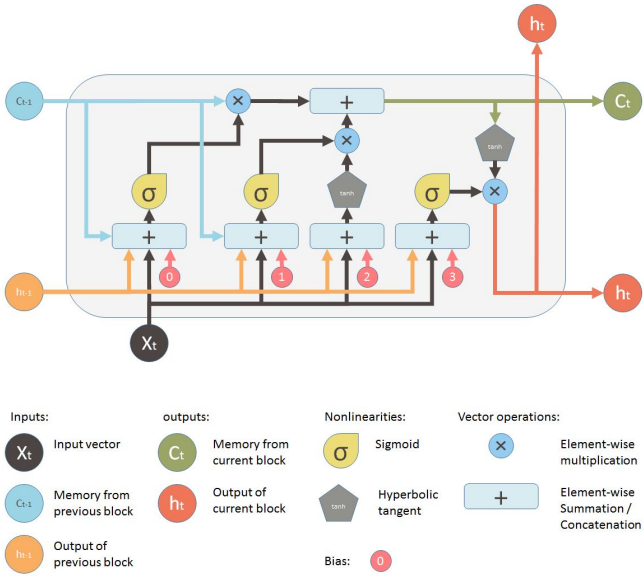

We have used a simple structure for LSTM RNN and the layers used in the architecture are shown below:

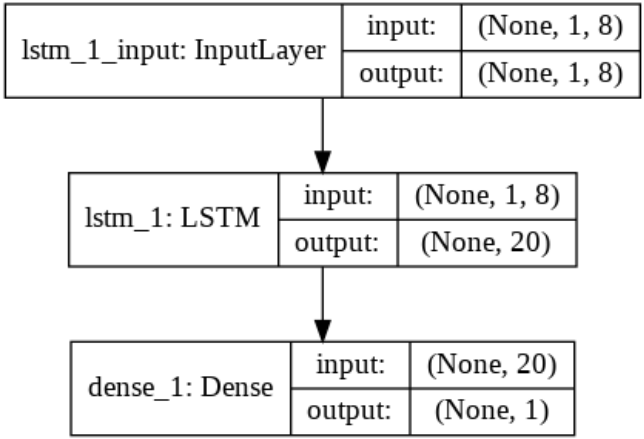

The model converged to the minimum loss in 67 epochs. The training cycle is presented in the figure. The model does not suffer from the problem of overfitting which is evident by the fact that the values of training loss and test loss are very close after convergence.

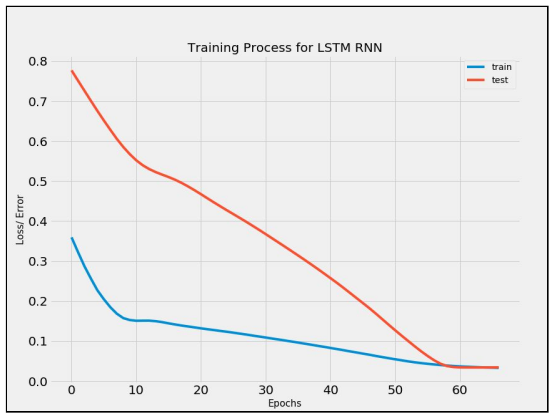

## IV. RESULTS

### A. Models' Performance

To measure the performance of each model, we have used 5 metrics of performance which are as follows:

- *Mean Absolute Percentage Error (MAPE):*

$$MAPE = \frac{1}{n} \sum_{t=1}^{n} \left| \frac{\hat{y} - y}{y} \right|$$

- *Mean Percentage Error (MPE):*

$$MPE = \frac{1}{n} \sum_{t=1}^{n} \left[ \frac{\hat{y} - y}{y} \right]$$

- *Root Mean Square Error (RMSE):*

$$RMSE = \left[ \frac{\sum_{t=1}^{n} (\hat{y} - y)^2}{n} \right]^{\frac{1}{2}}$$

- *Accuracy:*

$$Accuracy = 100\% - MAPE \times 100\%$$

All the models were trained on 90% of the available data and tested on the rest 10%.

| Train- Test Split | |
|---|---|
| **Train Data** | April 1994 - July 2016 |
| **Test Data** | August 2016 - December 2018 |

The following figure shows the predictions made by each of the models for the given testing time period. Intuitively, it is evident that the LSTM RNN model delivers the best and closest predictions followed by the SVM model and finally the VAR model. Quantitatively, the below table portrays the performance of each model the selected performance metrics. Clearly, the Root Mean Square Error (RMSE) is lowest in the case of the LSTM RNN model and highest in the case of the VAR model. The accuracy achieved by these models follows the same pattern of performance as the RMSE.

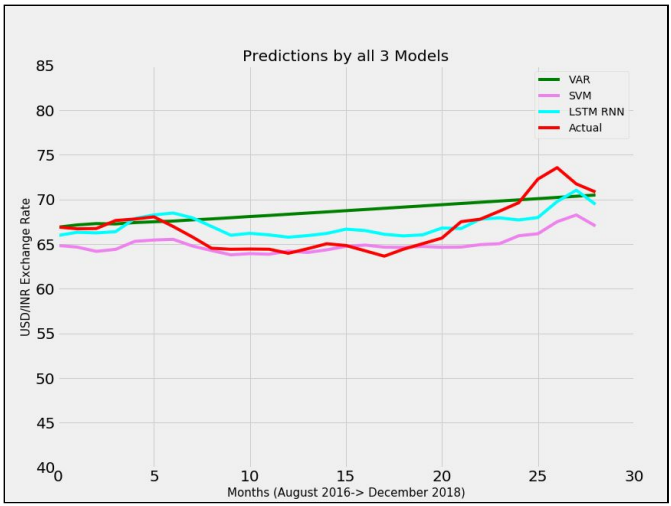

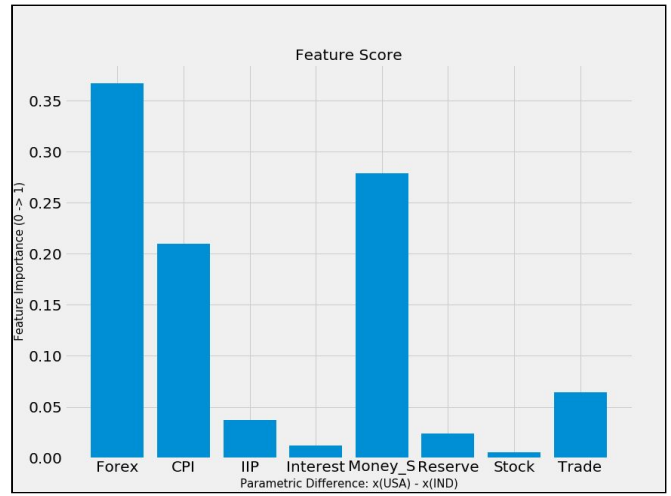

| Performance Metric | VAR Model | SVM Model | LSTM RNN Model |
|---|---|---|---|
| MAPE | 0.0369 | 0.0283 | 0.0217 |
| MPE | 0.0286 | -0.0261 | 0.0031 |
| RMSE | 2.9381 | 2.5585 | 1.6872 |
| Accuracy (%) | 96.31% | 97.17% | 97.83% |

B. *Feature Analysis*

- *Feature Importance Test:*

To view the effect of each feature in our analysis on the foreign exchange rate, we conducted a feature importance test using an ensemble of decision trees [32]. An ensemble of decision trees can compute the relative importance of each attribute by measuring the amount of variance that each feature decreases. The more a feature decreases the variance, the more important the feature is. As shown in the figure, the value of foreign exchange (forex) at time *t-1* has the highest feature importance for predicting the forex at time *t* followed by the value of the difference of CPIs of the two countries at time *t-1* and least dependent on the value of the difference of stock indices in the two countries at time *t-1*.

- *Feature Correlation:*

We use the Pearson Correlation Coefficient to quantify the correlation between all the variables in our analysis. The Pearson Correlation Coefficient is calculated as follows:

$$\rho_{X,Y} = \frac{Covariance(X, Y)}{\sigma_X \sigma_Y}$$

$$\text{or, } \rho_{X,Y} = \frac{E[(X - \mu_X)(Y - \mu_Y)]}{\sigma_X \sigma_Y}$$

where E is the expectation, represents standard deviation and represents mean.

The value of the correlation coefficient lies in the range [-1,1] where negative values represent negative linear relation and positive values represent positive linear relation among the two variables. Correlation Coefficient having 0 value means that there does not exist any correlation between the variables.

As evident from the figure, forex has the highest correlation with the difference in CPI and the difference in money supply and the lowest correlation with the interest rates difference.

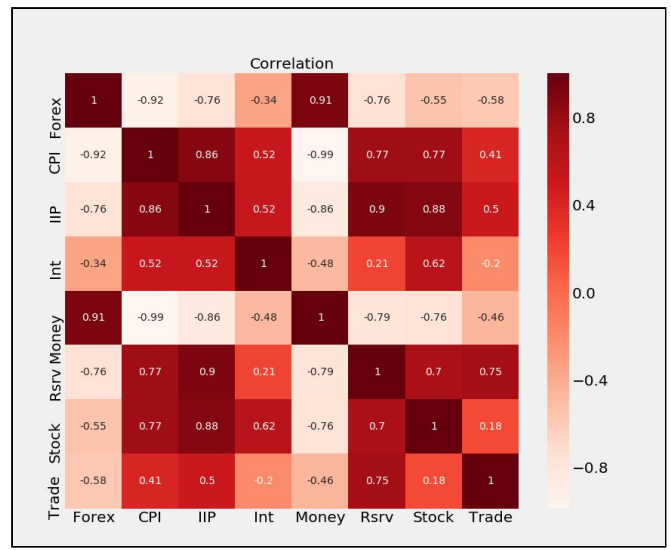

## Conclusions

A lot of information can be extracted from the changes in the values of macroeconomic variables which affect the foreign exchnage rates so as to make better predictions. Thus, multivariate time series analysis has shown commendable results while forecasting foreign exchange rates after including several other marcoeconomic indicators.

Models like ARIMA and VAR have set benchmark for economic and financial forecating due to their wide use in the lietrature, but modern machine learning and deep learning techniques clearly outperform the traditional econometric models. In our analysis, deep learning model, LSTM RNN, showed the best performnace while forecasting USD/INR foreign exchange rate followed by machine learning model, SVM, and then by traditional econometric model, VAR.